# *In Situ* Measurements of Stress Evolution in Silicon Thin Films During Electrochemical Lithiation and Delithiation

Vijay A. Sethuraman,[a,*] Michael J. Chon,[b] Maxwell Shimshak,[b] Venkat Srinivasan,[a] and Pradeep R. Guduru[b,*]

[a]Environmental Energy Technologies Division, Lawrence Berkeley National Laboratory
Berkeley, California 94720-8168, USA

[b]Division of Engineering, Brown University, Providence, Rhode Island 02912, USA

* Corresponding authors at: Division of Engineering, Brown University, 182 Hope Street, Providence, Rhode Island 02912, USA. Tel: +1 803 338 5742; Fax: +1 401 863 9009.
*Email address:* vj@cal.berkeley.edu (V. A. Sethuraman); Pradeep_Guduru@Brown.edu (P. R. Guduru)

We report *in situ* measurements of stress evolution in a silicon thin-film electrode during electrochemical lithiation and delithiation by using the Multi-beam Optical Sensor (MOS) technique. Upon lithiation, due to substrate constraint, the silicon electrode initially undergoes elastic deformation, resulting in rapid rise of compressive stress. The electrode begins to deform plastically at a compressive stress of *ca.* -1.75 GPa; subsequent lithiation results in continued plastic strain, dissipating mechanical energy. Upon delithiation, the electrode first undergoes elastic straining in the opposite direction, leading to a tensile stress of *ca.* 1 GPa; subsequently, it deforms plastically during the rest of delithiation. The plastic flow stress evolves continuously with lithium concentration. Thus, mechanical energy is dissipated in plastic deformation during both lithiation and delithiation, and it can be calculated from the stress measurements; we show that it is comparable to the polarization loss. Upon current interrupt, both the film stress and the electrode potential relax with similar time-constants, suggesting that stress contributes significantly to the chemical potential of lithiated-silicon.

**Keywords:** Silicon anode; lithium-ion battery; multi-beam optical sensor (MOS); mechanical dissipation; *in situ* stress measurement; open-circuit relaxation.

## 1. INTRODUCTION

Silicon based anodes are considered to be one of the most promising choice for the next generation of high energy density lithium-ion batteries due silicon's high charge capacity and relatively low density [1]. Since silicon undergoes huge volume expansion upon lithiation, which is responsible for cracking and capacity fading, measurement of stress evolution and mechanical dissipation is especially important [2]. *In situ* stress measurements in electrodes have been made in a wide variety of electrochemical systems in the past [3-11]. These measurements were based on a cantilever beam-bending method, in which curvature of the substrate is used to calculate stress in a film deposited on it through the Stoney equation [12,13]. Some of these studies [3,5,7] were conducted on electrode materials for lithium-ion batteries, primarily the cathode. However, none of them have gone beyond reporting stress evolution; and no attempt was made to calculate mechanical dissipation and to establish a quantitative connection between stress measurements and observed of mechanical damage. There appears to be just one reported measurement of stress evolution in silicon anodes, by Lee *et al.* [14], who employed the cantilever beam-deflection technique to observe stress evolution during lithiation and delithiation. However, their observations were only qualitative and did not attempt to convert beam-deflection potential signals into stress measurements.

In the cantilever beam-deflection method, a laser beam is reflected off the substrate surface; when the substrate acquires a curvature due to film stress, the position of the reflected beam on a photo-detector changes. By measuring the translation of the beam position, the substrate curvature is calculated. Recently, Láng *et al.* [15] carried out an analysis of this method, and showed that severe errors in stress measurement can result, if the refractive indices and incident angles of all media through which the laser beam travels are not accounted for, or not known accurately. Moreover, the technique is vibration sensitive, reducing the signal-noise ratio. Alternatively, the Multi-beam Optical Sensor (MOS) technique circumvents such issues by employing an array of parallel laser beams, and measuring the relative change in the spacing between them. MOS is insensitive to vibrations, and it has been used for thin-film stress monitoring in a variety of problems [16-18].

The objectives of this communication are to demonstrate real-time measurements of stress evolution in a silicon thin-film electrode during electrochemical lithiation and delithiation through the MOS technique; to demonstrate that such stress measurements can be used to calculate mechanical dissipation in the electrode during lithiation/delithiation, and compare it to polarization losses.

## 2. EXPERIMENTAL

*2. 1. Electrochemical cell*

Oxide free silicon wafers [50.8 mm diameter, 500 μm thick, (111)] were used as substrates for electrode fabrication. Silicon thin films were prepared by RF-magnetron sputtering (Edwards Auto 306 DC and RF Sputter Coater) of a silicon target (3" diameter disc, 99.995% Si, Plasmaterials Inc., Livermore, CA) at 200 W power and at a pressure of 0.667 Pa of Argon (99.995%). Previous studies have shown that silicon sputtered under these conditions result in amorphous films [19].Copper thin films were prepared by DC-sputtering of copper target (3" disc, 99.995%, Super Conductor Materials Inc., Suffern, NY) at 100 W and a pressure of 0.013 Pa of Argon. A 1500 nm copper thin film (*i.e.,* a Cu underlayer) was first sputtered onto the unpolished side of the Si wafer followed by the deposition of 250 nm silicon film. Previous studies show that the Cu underlayer is critical to the cycling of Si thin films [20]. Further, the Cu underlayer serves as a current collector, and aids in uniform current distribution on the Si electrode, an important role during the first-cycle lithiation. The electrochemical cell assembly is shown schematically in Figure 1.

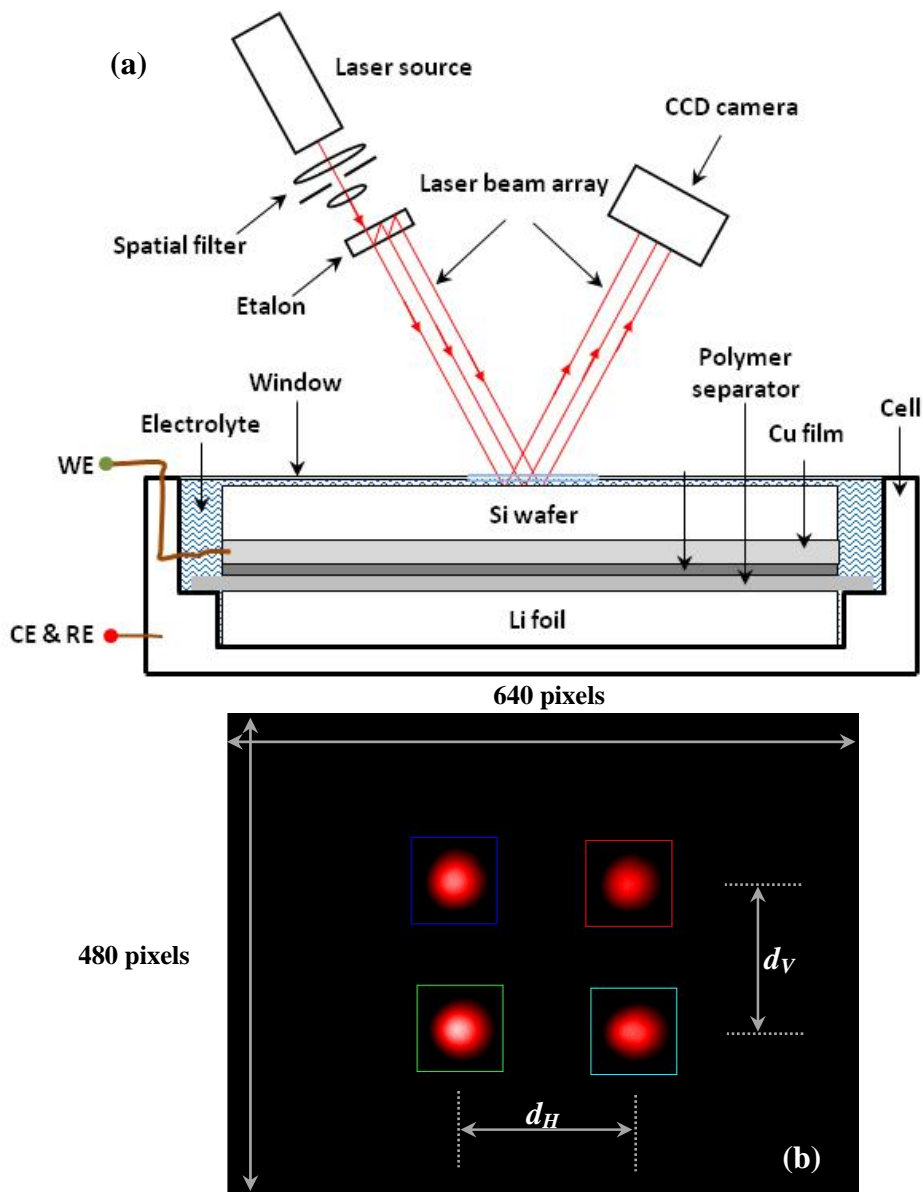

*Figure 1: (a) Schematic illustration of the electrochemical-cell assembly; and the MOSS setup to measure substrate curvature. (b) The relative change in the laser-spot spacing on the CCD camera's sensor is proportional to the curvature of the Si wafer. Note that the schematic is not drawn to scale.*

The Si wafer (coated with the Cu and Si thin films) was then assembled into a home-made electrochemical cell (Figure 1) with a lithium metal counter and reference electrode (diameter = 5.08 cm, thickness = 1.5 mm), and a woven Celgard 2500 separator (diameter = 5.2 cm, thickness = 50 µm, Celgard Inc., Charlotte, North Carolina). 1.2 M lithium hexafluoro phosphate in 1:2 (vol. %) ethylene carbonate:diethyl carbonate with 10% fluoroethylene carbonate additive (Novolyte Technologies Inc., Independence, OH) was used as the electrolyte.

*2. 2. Electrochemical measurements*

Electrochemical measurements were conducted in an environmental chamber (Labmaster SP, MBraun Inc., Stratham, New Hampshire) in ultra-high pure Argon atmosphere at 25°C (±1°C) using a Solartron 1470E MultiStat system (Solartron Analytical, Oak Ridge, TN), and data acquisition was done using CellTest System (Solartron Analytical). The cell was cycled galvanostatically at a current of 25 µA/cm$^2$ (*ca.* C/4 rate) between 1.2 and 0.01 V *vs.* Li/Li$^+$. The data acquisition rate was 1 Hz for all the electrochemical experiments. The lower limit of 0.01 V *vs.* Li/Li$^+$ was chosen to avoid lithium plating and also to avoid the formation of the crystalline Li$_{15}$S$_4$ phase. Two open-circuit relaxation experiments were conducted at approximately 50% state-of-charge, one from the lithiation side and another from the delithiation side. The input impedance of the instrument was 12 GΩ, and hence the current due to the open-circuit potential measurement was negligible.

*2. 3. In situ stress measurements*

Stress in the silicon thin-film was measured by monitoring the substrate curvature change during electrochemical lithiation and delithiation. The relationship between the biaxial film stress, $\sigma_f$, and the substrate curvature, $\kappa$, is given by the Stoney equation,

$$\sigma_f = \frac{E_s h_s^2 \kappa}{6 h_f (1-\upsilon_s)} \qquad 1$$

where $E_s$ and $\upsilon_s$ are the Young's modulus and the Poisson's ratio of the substrate respectively, and $h_s$ is the substrate thickness. The film thickness, $h_f$, is a function of state-of-charge and is expressed as $h_f = h_f^0 (1+2.7z)$, where z is the state of charge and $h_f^0$ is the intial film thickness. This expression takes into account 370% volume expansion corresponding to the maximum possible capacity of 3579 mAh/g for the lithiated-silicon system [21,22], and assumes one-dimensional volume expansion – *i.e.*, only the height of the thin-film electrode changes upon lithiation/delithiation. This is a reasonable assumption, because the film is constrained by the substrate from expanding in all directions except in the thickness direction. Also, small elastic-volume changes due to stresses are neglected in writing the above expression.

Substrate curvature was monitored with a MOS wafer curvature system (kSA-MOS, K-Space Associates, Inc., Dexter, Michigan), which is illustrated schematically in Figure 1. A 2x2 array of laser spots was used to measure curvature change in two orthogonal directions, a snapshot of which is shown in Fig. 1(b). Wafer curvature is calculated from,

$$\kappa = \frac{(d-d^0)}{d^0}\frac{1}{A_m} \qquad 2$$

where d is the distance between two adjacent laser spots on the CCD camera (see Figure 1(b), $d_H$ and $d_V$ are respectively the horizontal and vertical distances between the laser spots), $d^0$ is the initial distance, and $A_m$ is the mirror constant, given by $2L/\cos(\theta)$; L is the optical path length of the laser beam and $\theta$ is the incident angle of the beam on the substrate. The mirror constant is measured by placing a reference mirror of known curvature in the sample plane and measuring the relative change is the spot spacing. The vertical and the horizontal displacements of the laser spots (as a function of time) were recorded during all the electrochemical experiments on the silicon film at an acquisition rate of 1 Hz.

## 3. RESULTS AND DISCUSSION

Data reported in this communication are those obtained subsequent to the formation cycle of silicon. Since the sputtered films are already amorphous in nature [19], amorphization due to initial lithiation [23,24] does not occur, and the data corresponding to cycle 1 is similar to those obtained in subsequent cycles; however the duration of the first cycle is *ca.* 20% longer than the subsequent cycles, due to SEI layer formation and other losses associated with the first cycle lithiation and delithiation process. Figure 2 shows the cell potential and the film stress plotted against the capacity of the silicon thin-film electrode. In the Stoney equation, Young's modulus and Poisson's ratio corresponding to Si (111) plane were used [25,26]. The small difference in measured curvatures in the two perpendicular directions (*i.e.,* the X and Y curves in Figure 2b) is possibly due to variation in the deposited film thickness and/or slightly non-uniform current density across the Si surface.

Upon lithiation, the substrate prevents the in-plane expansion of the film, which results in compressive stress in the film, and it increases linearly with time (or capacity). If the diffusivity of Li in Si is taken to be $10^{-9}$ cm$^2$/s [27], the characteristic time for diffusion through the film thickness of 250 nm is only a fraction of a second. Since the experiments are conducted at C/4 rate, it is reasonable to assume that the Li concentration, and hence the in-plane stress are uniform through the film thickness. Moreover, if we assume that the strain induced by Li in Si is proportional to its concentration, then the linear increase in the compressive stress indicates elastic response. At compressive stress of about 1.7 GPa, the film appears to reach the elastic limit (which corresponds to a capacity of *ca.* 325 mAh/g), and begins to flow with further lithiation, in order to accommodate the additional volume expansion. The flow stress is seen to decrease with lithiation, reaching a value of about 1 GPa at a capacity of *ca.* 1875 mAh/g, at the cut-off potential. Hence, it can be concluded that the flow stress of lithiated Si decreases as the Li concentration increases (Note that, Figure 2b plots the true film stress, obtained by using the current value of the film thickness, which depends on the state of charge, in the Stoney equation. In order to evaluate the evolution of the film thickness, it is assumed that the strain in Si increases linearly with the state of charge and the volume expansion is 370% when the atom ratio of Li:Si is 3.75).

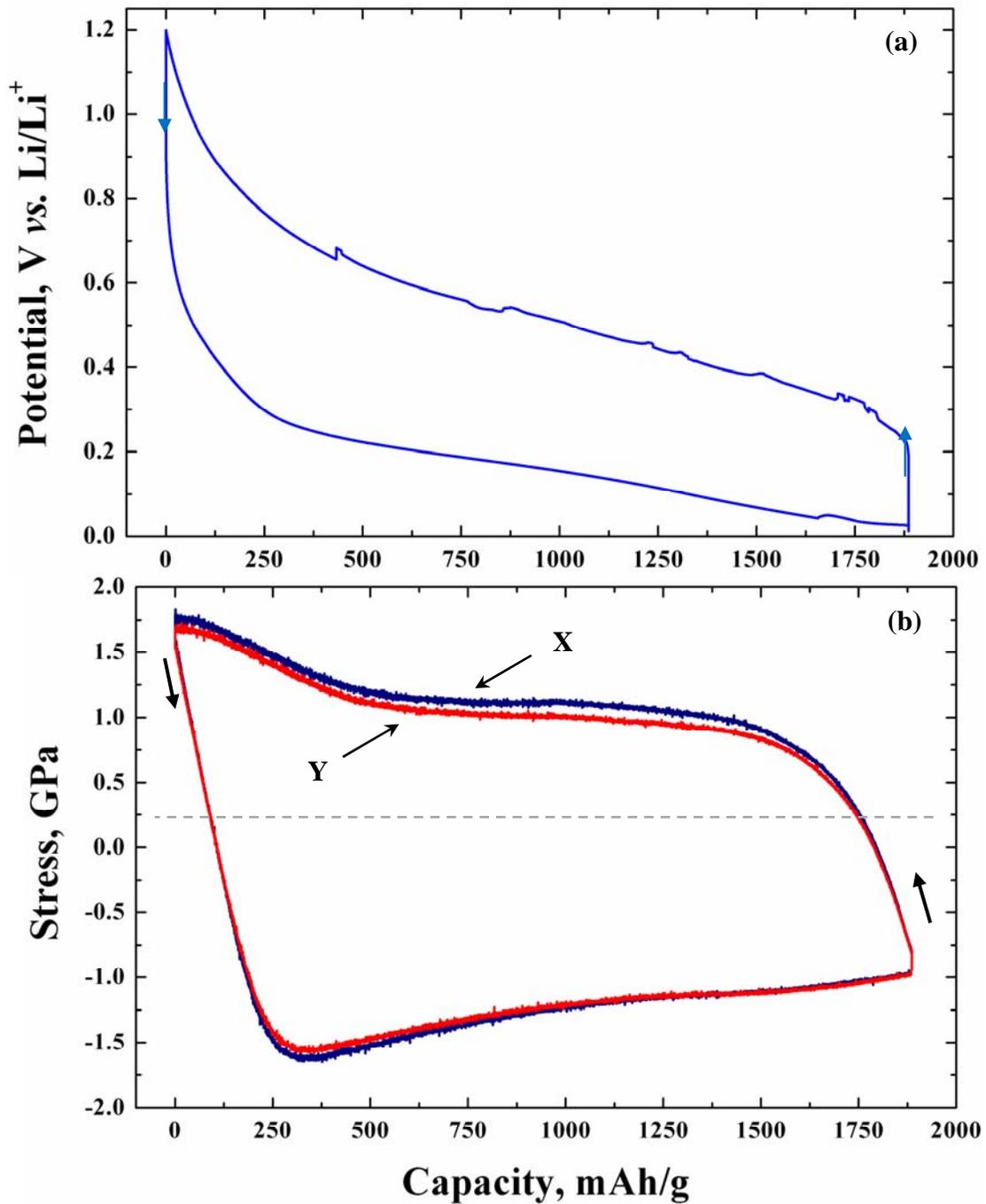

*Figure 2: (a) .Cell potential versus capacity curve corresponding to lithiation and delithiation of magnetron-sputtered amorphous Si thin-film electrode cycled at C/4 rate between 1.2 and 0.01 V vs. Li/Li$^+$, and (b) the corresponding stress calculated from the substrate-curvature data using the Stoney equation. The curves labeled X and Y correspond to the stress calculated from*

*the averaged horizontal and the vertical displacement of the spots, respectively. The arrows in both figures indicate cycling direction.*

Upon delithiation, the unloading is initially elastic; the stress reverses elastically until it becomes *ca.* 1 GPa in tension, where the film begins to flow in tension in order to accommodate the reduction in volume. The flow stress increases to about 1.75 GPa when the upper potential limit of 1.2 V is reached. Note that the stress response is similar in compression and tension, *i.e.,* at any state of charge, the flow stress in compression and tension are approximately the same. Thus, the film undergoes repeated compressive and tensile plastic flow during successive lithiation-delithiation processes, respectively. Since plastic flow dissipates mechanical energy, during delithiation, some of the electrical work done on the half-cell by the potentiostat is dissipated as plastic work in the silicon electrode. Similarly, during lithiation, some of the stored energy of the half-cell is dissipated in compressive plastic flow of the electrode. Thus, in addition to other polarization losses, plastic dissipation in the Si electrode must be taken into account. The measurements shown in Figure 2 correspond to the third charge-discharge cycle; subsequent cycles show a very similar behavior. Since the state of stress during lithiation and delithiation is different even at an identical state of charge, all material properties that depend on stress will be different during lithiation and delithiation.

The mechanical dissipation in the Si electrode can be calculated from the real-time stress measurements. In order to compare it with other polarization losses, consider the balance of energy during lithiation,

$$W_\mu = W_c^l + W_m^l + W_p^l \qquad 3$$

Where $W_\mu$ is the available energy in the half cell, $W_c^l$ is the work done by the cell on the potentiostat, $W_m^l$ is the mechanical dissipation in the Si electrode during compressive plastic flow, and $W_p^l$ is the sum of all other dissipations due to various polarizations (*i.e.,* kinetic, ohmic and transport). The first quantity $W_c^l$, is given by,

$$W_c^l = I \int_0^{t_l} V d\tau \qquad 4$$

where $t_l$ is the lithiation time, V is the cell potential, and I is the constant current at which lithiation is carried out. $W_c^l$ is calculated to be 1.77 J. The mechanical dissipation, $W_m^l$ is given as,

$$W_m^l = 2\upsilon \int \sigma_f d\dot\varepsilon \qquad 5$$

where $\upsilon$ is the film volume, $\sigma_f$ is the film stress, and $\dot\varepsilon$ is the in-plane plastic strain rate in the film during lithiation. Equation 5 can be re-written as,

$$W_m^l = 2A \frac{c_{max}}{t_f} \frac{d\varepsilon}{dc} \int_{t_p}^{t_l} \sigma_f h_f d\tau \qquad 6$$

where $h_f$ is film thickness, A is the total film area, c is the stoichiometric ratio of lithium to silicon, $c_{max}$ is its value at the end of lithiation, $t_p$ is the time at which plastic flow commences, and ($d\varepsilon/dc$) is calculated from the data reported by Obrovac et al. [28]. Although the values reported by Obrovac et al. are for crystalline Si-Li alloy phases, they are assumed to be valid for the amorphous Si-Li phase (which is the expected phase in the experiments reported here) as well. Amorphous thin films are known to undergo reversible shape and volume changes [29]. Further, the strain is assumed to depend linearly on concentration. Moreover, correction to equation 6 from elastic strain change during plastic flow is not considered because it is relatively small. In view of these assumptions, equation 6 should be viewed as a first order approximation for the plastic dissipation, rather than an exact calculation. Parameters used in the analyses are given in Table 1, from which $W_m^l$ is estimated to be 0.64 J.

A similar analysis can be done for delithiation as well, for which the work done by the potentiostat on the cell, $W_c^d$, is given by,

$$W_c^d = W_\mu + W_m^d + W_p^d \qquad 7$$

where all terms stand for the same quantities as in equation 3, except that the superscript d now stands for delithiation. Following a similar procedure, $W_c^d$ and $W_m^d$ are calculated to be 4.4 J and 0.48 J, respectively. It is reasonable to assume that the polarization losses are equal during lithiation and delithiation, i.e. $W_p^l = W_p^d$. Since the charge-discharge behavior is very similar in any two successive cycles (following the second cycle), the term $W_\mu$ can be taken to be equal in Eq. 3 and Eq.7. Eliminating $W_\mu$ between equations 3 and 7 gives a value of 0.75 J for the polarization loss $W_p^l$. Hence, the mechanical dissipation in a Si thin film electrode is comparable to the polarization losses elsewhere is the cell. An additional consequence of this result is that the contribution of stress to the chemical potential of lithiated silicon, and hence its potential, is significant and should be taken in to account for accurate cell modeling. Moreover, the foregoing experimental results and analysis suggest that new Si anode designs to minimize or eliminate the plastic dissipation can significantly improve the energy efficiency of a Si anode battery.

In order to explore the influence of stress on the silicon electrode potential, we carried out interruption of galvanostatic lithiation (or delithiation) as shown in Figure 3, and the subsequent potential and stress relaxations were monitored. The relaxations of potential and stress, upon current interruption during lithiation, are represented by PQ and AB, respectively. Upon the resumption of the galvanostatic experiment, the electrode potential and stress rapidly return to the potential and stress state (R and C), which is the same as that at the beginning of the interruption (P and A).

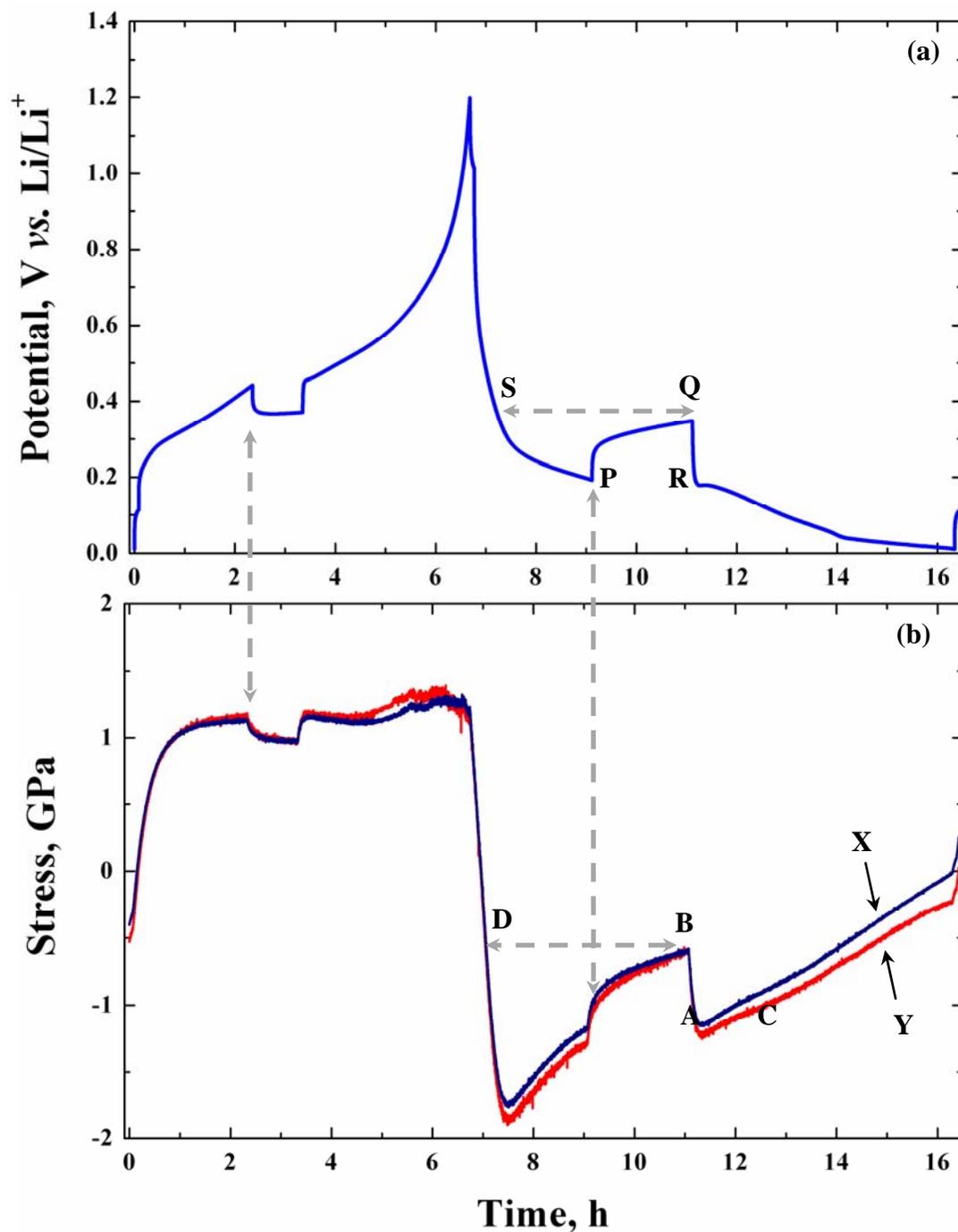

*Figure 3: Transient cell potential (a), and film stress calculated from the substrate-curvature data (b) is shown during lithiation, delithiation, and open-circuit-potential relaxation from both lithiation and delithiation directions. The OCP-relaxation experiments were conducted for 1 and 2 hours from the delithiation and lithiation sides, respectively, while the stress-changes were monitored. The curves labeled X and Y correspond to the stress calculated from the averaged horizontal and vertical displacement of the spots, respectively.*

Qualitatively, the potential and stress relaxations are similar; there is a rapid initial change followed by a slow decay. Although the time constants appear to be the same, the magnitude of relaxation depended on whether the current-interruption occurred during lithiation or delithiation.

**Table 1: Parameters used for the stress and energy analyses presented in this study.**

| Parameter | Definition | Value | Comments |
|---|---|---|---|
| $d_f$ | Film diameter (also Si wafer diameter) | 5.08 cm | Measured |
| $E_s$ | Young's modulus of Si (111) wafer | 169 GPa | Ref. 25 |
| $c_{max}$ | Maximum stoichiometric ratio in lithiated silicon system | 3.75 | Ref. 28 |
| $(d\varepsilon/dc)$ | Linear strain of lithiated silicon | 0.146 | Ref. 28 |
| $h_f^0$ | Initial film thickness | 250 nm | Measured |
| $h_s$ | Wafer thickness | 500 µm | Measured |
| $L/\cos(\theta)$ | Mirror constant | 2.7 m | Measured |
| $\upsilon_s$ | Poisson's ratio of Si (111) | 0.26 | Ref. 25 |
| $\rho_f$ | Film density | 2.3 g/cm$^3$ | Ref. 25 |

The potential and stress relaxations can be partly explained by the small concentration change associated with the relaxation of the double-layer, which drives both the electrolyte-reduction reaction, and the delithiation reaction (when the system is interrupted from the lithiation side, and vice versa). However, concentration change due to double-layer relaxation alone cannot explain the large stress drop and potential change. For example, from Figure 3b, if the stress drop during lithiation is attributed to concentration change alone, then the double layer would have to delithiate the electrode almost completely to the value at D. In which case, upon current resumption, it should have taken about 2 hours to return to the state represented by A and P (*i.e.,* the duration for DA and SP). However, the pre-interrupt stress and potential states are regained rapidly in just a few minutes (BC and QR in Fig. 3). Hence, alternative mechanisms for stress and potential drop need to be sought. A potential explanation is that the drop in stress magnitude is driven by the viscoplastic relaxation mechanisms in the amorphous Si-Li alloy; since the mechanical dissipation analysis above establishes a coupling between stress and potential, viscoplastic stress relaxation would be associated with a corresponding potential relaxation. Although this is only a plausible explanation, the experimental capabilities demonstrated in this communication will allow us to investigate it systematically in near future.

## 4. CONCLUSIONS AND FUTURE WORK

We have demonstrated the use of the multi-beam optical sensor technique to measure stress evolution in a silicon thin film electrode during lithiation and delithiation; and during the open-circuit relaxation upon current-interrupt during lithiation or delithiation. Stress evolution upon electrochemical cycling reveals that the Si thin-film electrode undergoes repeated cycles of compressive- and tensile-plastic flow, dissipating mechanical energy. The stress evolution data enables estimation of the mechanical dissipation, which was found to be comparable to the polarization losses elsewhere in the cell. This observation also suggests that stress contributes significantly to the chemical potential of lithiated silicon and hence the electrode potential. Further experiments aimed at understanding the influence of the mechanical stresses on the equilibrium potential of the lithiated-silicon, as well as the stress evolution during the amorphization of a crystalline-silicon electrode during initial lithiation are currently ongoing in our laboratory. Such experiments are expected to provide insights to understand the potential hysteresis and to suggest ways to reduce hysteresis, which can increase the energy efficiency of the cell.

## 5. ACKNOWLEDGEMENTS

Authors at the Lawrence Berkeley National Laboratory gratefully acknowledge the support by the Assistant Secretary for Energy Efficiency and Renewable Energy, Office of Vehicle Technologies, the United States Department of Energy, under contract no. DE-AC02-05CH11231. Authors at Brown University gratefully acknowledge the support by the Materials Research, Science and Engineering Center (MRSEC) sponsored by the United States National Science Foundation, under contract no. DMR0520651. Helpful discussions with Professor Allan Bower (Brown University) are gratefully acknowledged.